\documentclass{aastex}    
\usepackage{emulateapj5}  
\usepackage{psfig}
\def\aa{A\&A}
\def\del#1{}
\def\RM{{\rm RM}}
\def\PA{{\rm PA}}
\def\vx{\vec{x}}
\def\lesssim{\mathrel{\hbox{\rlap{\hbox{\lower4pt\hbox{$\sim$}}}\hbox{$<$}}}}
\def\gtrsim{\mathrel{\hbox{\rlap{\hbox{\lower4pt\hbox{$\sim$}}}\hbox{$>$}}}}

\begin{document}

\title{Are the Faraday Rotating Magnetic Fields Local to Intracluster
Radio Galaxies?}

\shorttitle{Are the Faraday Rotating Magnetic Fields Local to Intracluster
Radio Galaxies?}

\author{Torsten A. En{\ss}lin\altaffilmark{1}, Corina
Vogt\altaffilmark{1}, T. E. Clarke\altaffilmark{2,3}, Greg
B. Taylor\altaffilmark{3}}

\altaffiltext{1}{Max-Planck-Institut f{\"u}r Astrophysik,
Karl-Schwarzschild-Str.1, 85740 Garching, Germany}
\altaffiltext{2}{Department of Astronomy, University of Virginia,
P.O. Box 3818, Charlottesville, VA 22903-0818}
\altaffiltext{3}{National Radio Astronomy Observatory,
Socorro, NM 87801, USA}

\begin{abstract}
We investigate the origin of the high Faraday rotation measures (RMs)
found for polarized radio galaxies in clusters.  The two most likely
origins are, magnetic fields local to the source, or magnetic fields
located in the foreground intra-cluster medium (ICM). The latter is
identified as the null hypothesis.  Rudnick \& Blundell (2003) have
recently suggested that the presence of magnetic fields local to the
source may be revealed in correlations of the position angles (PAs) of
the source intrinsic linear polarization and the RMs.  We investigate
the claim of Rudnick \& Blundell to have found a relationship between
the intrinsic PA$_0$ of the radio source PKS~1246-410 and its RM, by
testing the clustering strength of the PA$_0$-RM scatter plot.  We
show that the claimed relationship is an artifact of an improperly
performed null-experiment.
We describe a gradient alignment statistic aimed at finding 
correlations between PA$_0$ and RM maps. This statistic does not
require any null-experiment since it gives a unique (zero) result in
the case of uncorrelated maps. We apply it to a number of extended
radio sources in galaxy clusters (PKS~1246-410, Cygnus~A, Hydra~A, and
3C465). In no case is a significant large-scale alignment of PA$_0$ and
RM maps detected. We find significant small-scale
co-alignment in all cases, but we are able to fully identify this with
map making artifacts through a suitable statistical test.  We conclude that
there is presently no existing evidence for Faraday rotation local to radio
lobes. Given the existing independent pieces of evidence, we favor the null
hypothesis that the observed Faraday screens are produced by intracluster
magnetic fields.
\end{abstract}

\keywords{galaxies: clusters: general --- magnetic fields ---
polarization --- radio continuum: galaxies}

\maketitle

\section{Introduction}

A detailed summary of the observational evidence for the presence of
magnetic fields embedded within the thermal gas in clusters of
galaxies is presented in two recent review papers (Carilli \& Taylor
2002, and Widrow 2002) and is only briefly summarized here. 

Magnetic fields in galaxy clusters are known to exist due to the
detection of diffuse cluster wide synchrotron emission (Willson 1970)
in a number of clusters. This emission is not associated with
individual galaxies in the cluster and is observationally classified
as {\it halo} or {\it relic} emission (Feretti 1999). Although
initially thought to be relatively rare objects, recent surveys have
significantly increased the number of radio halos and relics
(Giovannini, Tordi, \& Feretti 1999; Giovannini \& Feretti 2000;
Kempner \& Sarazin 2001; Bacchi et al.\ 2003). The presence of this
diffuse synchrotron emission reveals the widespread distribution of
magnetic fields within the intracluster medium (ICM) in these
clusters. Equipartition assumptions provide minimum energy magnetic
field estimates in radio halos of 0.1 - 1 $\mu$G (Feretti 1999; Bacchi
et al.  2003) and 0.4 - 2.7 $\mu$G in radio relics (En{\ss}lin et al.\
1998).

Estimates of volume averaged intracluster magnetic field strengths can
be obtained by comparing the synchrotron and inverse Compton emission
(e.g., Harris \& Grindlay 1979, Rephaeli et al. 1987). The same
relativistic particle population which produces the diffuse
synchrotron emission is expected to up-scatter the ambient photon
field in the ICM to produce inverse Compton X-rays. Typical estimates
yield a volume-averaged intracluster magnetic field in the range of
0.2 - 1 $\mu$G (see e.g.\ Carilli \& Taylor 2002 and references
therein). We emphasize that inverse Compton based estimates have to be
regarded as lower limits.

Faraday rotation measure (RM) studies of extended radio sources
embedded in clusters probe the magnetic field component
integrated along the line-of-sight, weighted by the plasma electron
density. Faraday rotation is the rotation of the position angle (PA)
of the plane of linear polarization of radio waves, which traverse a
magnetized plasma. The RM is the proportionality constant of this
$\lambda^2$-dependent effect: $\PA(\lambda)= \PA_0 + \lambda^2\,\RM$,
where PA$_0$ is the source intrinsic PA, only directly observable at
shortest wavelength. The RM and PA$_0$ can be obtained by
multi-frequency measurements of a polarized radio source.

Carilli \& Taylor (2002) review the generally large RM values obtained
from extended radio sources embedded in clusters, and summarize the
evidence that indicates that the observed RM of embedded sources are
indeed probing a foreground ICM, with conservative estimates of
magnetic field strengths between a few $\mu$G and 10s of
$\mu$G. Similar magnetic field strengths are determined from
statistical RM studies by Kim, Tribble \& Kronberg (1991), and Clarke,
Kronberg \& B{\"o}hringer (2001).

Finally, the asymmetric depolarization of double radio lobes embedded
in galaxy clusters can be understood as resulting from a difference in
the Faraday depth of the two lobes (Laing 1988; Garrington\ et~al.\
1988) which strongly supports the association of RMs with ICM
magnetic fields. We note that such a difference in Faraday depth might
also be explained by local effects near the source, if there is a
difference in the mixing layers of the magnetized radio plasma with
the dense environmental gas between the radio lobe head side and the
back-flow side of an FR~II radio galaxy. In contrast, this scenario
would have difficulties to explain the observed strong RM and
depolarization asymmetry of FR~I radio galaxies like Hydra~A (Taylor
\& Perley 1993), which are not believed to have back-flows.  However,
this discussion shows that the association of the RM with the
intra-cluster medium -- in the following regarded as the null
hypothesis -- is not unambiguous, since it could also be produced in a
magnetized plasma skin or mixing layer of the observed radio galaxy
(Bicknell\ et~al.\ 1990). For that reason Rudnick \& Blundell (2003,
hereafter RB) used Faraday rotation measure observations of PKS
1246-410 by Taylor, Fabian \& Allen (2002) in an attempt to estimate
the fraction of the measured RM signal which is local to the radio
source.  If most of the RM signal is local to the radio source
then the derived ICM magnetic field strength should be significantly
lower than if all of the RM originates in the ICM.

In the following we examine RB's claim for evidence of source-local
magnetic fields. We do this first through a general argument
(Sect. \ref{sec:scatter}) followed by an illustrative numerical
example (Sect. \ref{sec:example}). Then, in order to have an unbiased
tool to measure cross correlations of RM and PA maps, we specify the
mathematical requirements of a suitable statistic
(Sect. \ref{sec:req}), and construct a {\it gradient alignment
statistic} which meets our quality criteria
(Sect. \ref{sec:align}). This statistic is also sensitive to
correlated noise resulting from the map making imperfections (Appendix
\ref{app:correlatednoise}). We therefore design a {\it gradient vector
product statistic}, which is only sensitive to such imperfections, in
order to separate spurious signals from any astrophysical signal
(Sect. \ref{sec:mapartefacts}).  Application of our approach to
several datasets (PKS~1246-410, Cygnus~A, Hydra~A, and 3C465) does not
reveal any evidence for source-local magnetic fields
(Sect. \ref{sec:results}).  We conclude (Sect. \ref{sec:summary}) that
the null hypothesis that the RMs are connected to the ICM is not only
consistent with, but also strongly favored by present data. The basic
assumption for Faraday rotation based ICM magnetic field estimates of
RM being mostly generated by the ICM magnetic fields seems therefore
to be valid.

\section{PA$_0$-RM scatter plots\label{sec:scatter}}

If the Faraday rotation observed in cluster radio galaxies is produced by
a mixing layer between the radio lobe and the ICM gas then it is possible
that the magnetic structures within the Faraday screen are somehow related
to the magnetic field orientation within the radio lobes. In such a case,
co-spatial structures in RM and PA$_0$ maps are possible since both
quantities contain geometrical information about the source local magnetic
field geometry: The PA$_0$ gives the plane-of-the-sky direction
perpendicular to the sky-projected magnetic field within the radio lobe,
and the RM the line-of-sight component of the magnetic field
within the Faraday rotating medium.

The idea of RB is to test the above scenario by searching for
the expected co-spatial structures in the PA$_0$ and RM maps of the
radio galaxy PKS~1246-410.  One difficulty with such an analysis is
that no direct correlation between the PA$_0$ and RM values can be
expected even for co-spatial structures, since there is no generic
reference point for PA$_0$s.  Therefore RB analyzed the
PA$_0$-RM scatter plot generated by plotting for each map pixel the
PA$_0$ and the RM value in the same diagram. They argue that local
co-alignment of PA$_0$ and RM values should lead to a strongly
clustered point distribution in the scatter plot. Since statistically
independent PA$_0$ and RM maps may also produce such clusters they
perform a null-experiment. They generate synthetic RM maps, which have
the same power-spectrum as the observed RM map, but random
phases. With these simulated maps they repeat their analysis and find
much less clustering in the scatter plot. From this they conclude that
there is a strong co-alignment in the PA$_0$-RM maps of PKS~1246-410,
and therefore most of the RM should be local to the radio source. In
order to give further support to their claim, they report a number of
regions in a larger sample of radio galaxy maps where they see
corresponding structures in the RM and PA$_0$ maps.

This latter finding is statistically very questionable, since within
any two large datasets the human eye often finds apparent correlations
within some sub-regions, even if the two datasets are
uncorrelated. Furthermore, observed RM and PA$_0$ always carry
correlated noise, as recognized by RB and analytically demonstrated in
Appendix \ref{app:correlatednoise}, since they are generated from the
same set of PA maps. In the worst case the noise produces
step-function like artifacts at the same location in RM and PA$_0$
maps. We note that correlated steps are present within the aforementioned
regions of RM and PA$_0$ maps and give -- at least to our eyes -- the
dominant contribution to the PA$_0$-RM correlation impression reported
by RB.

However, the claim of a statistical detection of a PA$_0$-RM
correlation can be rigorously investigated.  The
null-experiment applied by RB is designed to have the same two-point
autocorrelation function as the observational data, but higher order
correlations are neglected by the random phase realization.
Unfortunately, the chosen statistic is very sensitive to such higher
order correlations. Any clustering in the PA$_0$-RM scatter plot is
caused by having patches of nearly constant values in the PA$_0$ and
RM maps. The appearance of patches requires special relations of the
Fourier phases, or -- equivalently -- non-trivial higher order
correlation functions to be specified.

\begin{figure}
\epsscale{1.0} 
\plotone{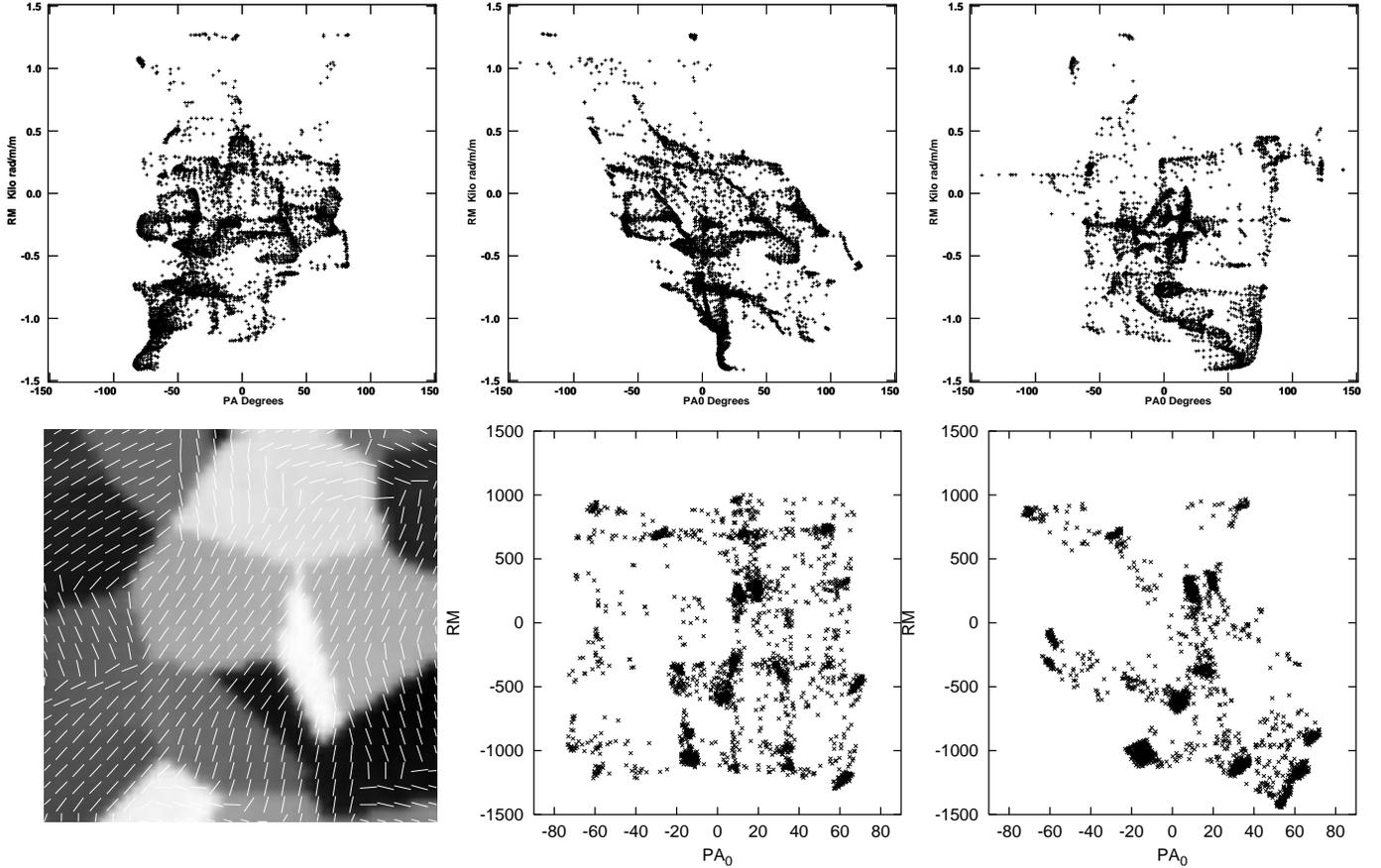} 
\figcaption{\label{fig:pks1}\label{fig:scatter_obs}\label{fig:scatter}
{\bf Upper left panel:} Uncorrected PA-RM scatter plot of PKS~1246-410
(similar to Figure 4 of RB). The underlying maps were smoothed by a
median weight filter in the same way as RB smoothed the data in order
to amplify the clustering.
{\bf Upper middle panel}: Similar to the upper left panel, but now
corrected PA$_0$-RM scatter plot of PKS~1246-410. Note the appearance
of diagonal stripes, which are artificial correlations due to errors
in the RM maps, which imprint themselves onto the PA$_0$ in the
Faraday de-rotation step. If they were true correlations, no specific
direction within the scatter plot should be favored. The decreasing
trend from left to right indicates actually an anti-correlation, as
predicted in Appendix \ref{app:correlatednoise} as a signature of
observational noise which propagates through the map making process.
{\bf Upper right panel:} PA$_0$-RM scatter plot of PKS~1246-410, but
with the RM maps of the Eastern and Western radio lobe exchanged as a
null-experiment. The clustering remains despite the fact that PA$_0$
and RM maps are now independent by construction. 
{\bf Lower left panel:} Simulated PA$_0$ (white lines) and RM
(greyscale) maps with independent coherence patches.
{\bf Lower middle panel:} PA$_0$-RM scatter plot of the simulated
independent PA$_0$ and RM maps shown on the left.
{\bf Lower right panel:} PA$_0$-RM scatter plot of simulated maps with
co-aligned PA$_0$ and RM coherence patches. Note the horizontal and
vertical patches in the plot of independent PA$_0$ and RM. Similar
structures are clearly visible in the scatter plots of PKS 1246-410
but are absent in the scatter plots of co-aligned PA$_0$ and RM.
We note that these qualitative impressions about the bridges between
clusters in the scatter plot are reflected in the results of our two
gradient statistics introduced in Sects. \ref{sec:align} and
\ref{sec:mapartefacts}: The gradient vector product statistic $V$ is
sensitive to an imbalance between the diagonally elongated features
(increasing versus decreasing diagonals which correspond to correlated
versus anti-correlated changes in RM and PA$_0$). The gradient
alignment statistic $A$ measures if both variables RM and PA$_0$
change simultaneously as one moves in the maps from patch to patch
(and therefore in the scatter plot from one cluster to another, as in
the lower right panel), or if these changes happen sequentially
(e.g. first RM changes rapidly, and later PA$_0$ changes rapidly, as
in all upper and in the lower middle panels).}
\end{figure}

The strength of the clustering in the scatter plot is indeed strongest
if the PA$_0$ and RM structures are correlated. However, the
clustering does not disappear if the RM and PA$_0$ patches have
independent distributions, since every PA$_0$ patch is still overlaid
by a small number of RM patches, so that the associated clustering in
the scatter plot only gets split into a corresponding number of
smaller clusters. These clusters happen to be co-aligned on a vertical
constant PA$_0$ line in the plot, since they all belong to the same
PA$_0$ coherence patch.  A corresponding effect splits the pixels of
an RM cell into a horizontal line of clusters.  Such horizontal and
vertical structures are indeed visible in the scatter plot of
PKS~1246-410 (Fig. \ref{fig:scatter_obs}, also Fig. 4 of RB).

We note that a proper null-experiment, which would maintain the
higher-order correlations, can also be constructed empirically by
exchanging subregions of the observed RM image
from one radio lobe to the other. This should keep the same RM
correlation functions, but will destroy any real PA$_0$-RM alignment since
the different regions of the source are independent. We have performed
this lobe switching by simply dividing the source in roughly two equal
regions about the center and shifting the coordinates in right
ascension such that the two subregions overlap. We have then plotted
the PA$_0$ of the east (west) lobe against the RM of the west (east)
lobe. This source division is preferable to a reflection about the
declination axis through the core as our lobe shifting results in a
correlation of central source region PA$_0$ with the outer lobe region
of RM thus eliminating possible radial influences on the
correlation. The results of such an experiment are also presented in
Fig. ~\ref{fig:pks1}. Clustering of points in the PA$_0$-RM scatter
plots are clearly seen in both plots in contrast to RB's claim that
clustering results from a relation between PA$_0$ and RM.

While RB noted that higher order correlations in independent RM and PA$_0$
maps can systematically produce spurious signals in their statistic. It
seems to follow that this implies that their statistic does not allow any
conclusions about the existence of PA$_0$-RM co-alignment. They argue that the
presentation of a counter example, as we provide in the next section, does not
invalidate their approach, since it cannot be demonstrated that the example
used is realized in nature. However, we note that RBs approach will not be
useful if patchiness is important. Already an inspection by eye of the maps
of PKS~1246-410 in comparison to RB's random phase maps shows that nature (or
the map making software) produces patchy maps which must exhibit clustering in
PA$_0$-RM scatter plots, independent of whether PA$_0$-RM alignment is present
or not.

In the next section, we provide a patchy map generating algorithm,
which illustrates our general argument. More importantly, it allows us
to test the {\it gradient alignment statistic}, introduced in
Sect. \ref{sec:align}.

\section{Synthetic patchy maps\label{sec:example}}

For illustration purposes we construct a simple example, which captures the
main effects, but is not meant to be exact in all respects. We
construct patchy RM and PA$_0$ maps, which are statistically
independent from each other. For both maps we use a similar
recipe\footnote{The computer code can be requested from TAE ({\tt
ensslin@mpa-garching.mpg.de}).}.

First a number $N$ of random seed points $\vec{X}_i$ ($i\in\{1...N\}$)
within a square area is drawn, and then the area is split into cells
around the seed points by means of a Voronoi-tessellation: Each point
$\vx$ of the area belongs to the cell of its nearest seed $\vec{X}_i$.
Then each seed is attributed a random value $\psi_i$ ($\psi$ stands in
the following for both RM and PA$_0$) and a small two-dimensional
random vector $\vec{\nabla} \psi_i$ (= auxiliary RM or PA$_0$
gradients within the patches, only used for the map
construction). Each pixel $\vx$ within the cell of seed $i$ gets a
value
\begin{equation}
\psi(\vx) = \psi_i + \vec{\nabla} \psi_i \cdot (\vx
-\vec{X}_i) + \sigma (\vx)\,,
\end{equation}
where $\sigma(\vx)$ is a small random noise term. The resulting map
consists of patches with nearly constant values, but which exhibit
some internal trends and noise. The adopted parameters are
described in Appendix \ref{app:par}.

The PA$_0$ and RM maps are slightly smoothed, a $20\%$ border region
is cut away in order to suppress edge effects, and a PA$_0$-RM scatter
plot is generated. A typical realization of such a scatter plot is
shown in Fig. \ref{fig:scatter}. A strongly clustered distribution is
visible even though the individual PA$_0$ and RM maps were completely
independent. Furthermore, nearly horizontal and vertical
chains of clusters are visible for the reason given in
Sect. \ref{sec:scatter}.  The map smoothing produces bridges between
these clusters, since it gives intermediate values to pixels which are
at boundaries of PA$_0$ or RM cells.

The deviations from the strict horizontal and vertical directions
visible in the scatter plot of PKS~1246-410 should be caused by trends
within the coherence cells.  We note that such structures are also
visible in the simulated scatter plots of RB, although the smooth
realizations of their RM maps has smeared them out (see their Fig. 2
for a comparison of the patchiness of observed and simulated
RM maps).

We therefore conclude that the data of PKS~1246-410 favors
statistically independent PA$_0$ and RM maps.  The occurrence of
vertical and horizontal lines in the scatter plot of the observational
data demonstrates that the PA$_0$ and RM patches are indeed misaligned.

In order to further investigate this, we construct a model in which
the PA$_0$ and RM patch positions are absolutely identical. We
construct this by moving the Voronoi-tessellation seed points of the
RM map to the locations of nearby seed points of the PA$_0$ map,
thereby assuring that there is a one-to-one mapping. All other
variables (RM$_i$, PA$_{0,i}$ etc) were kept as before. The recomputed
RM map has therefore exactly the same patch locations as the PA$_0$
map.  The horizontal and vertical cluster alignments and stripes are
absent there (see Fig. \ref{fig:scatter}). There are now several
stripes with diagonal orientations due to pixels at the PA$_0$-RM cell
boundaries which received simultaneously intermediate values in PA$_0$
and RM by the smoothing.

\section{Requirements for suitable statistics\label{sec:req}}

In order to quantitatively investigate the location of the Faraday
rotation measure screen we must develop a statistical basis for
distinguishing local and foreground effects. Therefore, we have to
construct a statistic which allows us to search for local correlations
between $\PA_0$ and RM without requiring a global relation between
these quantities. In order to design a proper statistic one should
first specify mathematical requirements in order to avoid potential
pitfalls. We require any suitable statistic $A$ to fulfill the
following conditions:
\begin{enumerate}
\item \label{cond:nofunc}
The statistic should be sensitive to the presence of correlated
spatial changes of PA$_0$ and RM, independent of the local values of
these quantities. Correlations in PA$_0$ and RM maps can not be
expected to follow a functional form like $\RM(\vx)
=f(\PA_0(\vx))+\delta f(\vx)$.
\item \label{cond:nonull}
The statistic $A$ should not require a null experiment, which
is always problematic since the construction of a synthetic dataset
with all the important statistical properties of the real data
reproduced is very difficult. 
\item \label{cond:unique}
The statistic has to provide unique expectation values in case
of uncorrelated ($A=0$) and in case of fully correlated ($A= 1$) maps.
\item \label{cond:analytic}
The statistic should be analytic and sufficiently simple, allowing
that basic properties can be derived and understood analytically.
\item \label{cond:monotonic}
The significance of the statistic should increase
monotonically with the map size.
\end{enumerate}
The approach of RB to calculate the autocorrelation of PA$_0$-RM
scatter plots fulfills our requirement No. \ref{cond:nofunc}, or --
more exactly -- requirement No. \ref{cond:nofunc} is inspired by their
work. However, their approach requires a null experiment (violating
No. \ref{cond:nonull}), since it is not clear a-priori what the
meaning of the derived correlation in the scatter plot is (violating
No. \ref{cond:unique}).  Although it is possible to derive analytic
equations describing the autocorrelation of a scatter plot of two
quantities, RB's application of a median weight filter to the RM and
PA maps is a strongly non-linear operation which makes it very hard to
understand the mathematical properties of their data treatment
(violating No. \ref{cond:analytic}). Finally, if one imagines the map
size growing to infinity, one easily realizes that the scatter plot
will be smoothly filled with points, independent if there were local
correlations between PA$_0$-RM or not. Therefore the significance of
the method of RB starts to decrease when the clumps are so densely
spaced that they start to merge, and vanishes in the limit of an
infinitely large map (violating No. \ref{cond:monotonic}). Applying a
median weight filter to maps -- as RB did -- sharpens the clumps in
the scatter plot considerably, thus suppresses the clump merging and
therefore enhances the autocorrelation signal.

\section{Gradient alignment statistics\label{sec:align}}
In order to construct a statistic which fulfills our requirements, we
introduce the {\it gradient alignment statistic} $A$ of different maps
by comparing the gradient $\vec{p}= \vec{\nabla}\,$RM and
$\vec{q}=\vec{\nabla}\,$PA$_0$.\footnote{We calculate gradients of a
quantity $\psi$ defined on our maps by assigning the pixel position
$(i,j)$ with $\vec{\nabla} \psi = (\psi_{i+1,j} + \psi_{i+1,j+1} -
\psi_{i,j} - \psi_{i,j+1}, \psi_{i,j+1} + \psi_{i+1,j+1} - \psi_{i,j}
- \psi_{i+1,j})$, where $\psi_{i,j}$ denotes the value of $\psi$ at
the pixel position $(i,j)$. Neither the small diagonal shift by $1/2$
pixel in $i$ and $j$ directions, nor the missing normalization of the
so defined gradient have any effect on our statistics.
PA$_0$-gradients are calculated
using subtraction modulo $180^\circ$ in order to account for the
cyclic nature of PAs.
We note that we apply this scheme to the maps directly, even though
for our synthetic maps gradient-like auxiliary quantities were defined
and used during their construction.}
The idea is to check for alignment of $\vec{p}$ and $\vec{q}$,
indicating correlated changes in PA$_0$ and RM, thus fulfilling
condition No. \ref{cond:nofunc}. Since the absolute values of RM and
PA$_0$ are not of any significance for the question of co-alignment,
the comparison should give the same signal for parallel and
anti-parallel gradients. Therefore, instead of the scalar product
\begin{equation}
\vec{p} \cdot \vec{q} =  p\,q\,(\cos\phi_p\,
\cos\phi_q\,+ \sin\phi_p\, \sin\phi_q) =p_x\,q_x + p_y\,q_y 
\end{equation}
of $\vec{p} = (p_x, p_y) = p\,(\cos\phi_p, \sin\phi_p)$ and $\vec{q} =
(q_x, q_y) = q\,(\cos\phi_q, \sin\phi_q)$ we construct an {\it
alignment product}
\begin{eqnarray}
\langle \vec{p}, \vec{q} \rangle &=&  p\,q\,(\cos 2\phi_p\,
\cos2\phi_q\,+ \sin2\phi_p\, \sin2\phi_q)\nonumber\\
& =& \frac{(p_x^2 -p_y^2)(q_x^2 - q_y^2) + 4\,p_x\,p_y\,q_x\,q_y }{
\sqrt{p_x^2 + p_y^2} \sqrt{q_x^2 + q_y^2}}.
\end{eqnarray}
The alignment product has the properties $\langle \vec{p},
\alpha\,\vec{p} \rangle = |\alpha|\, p^2$ for $\alpha$ being any real
(positive or negative) number, and $\langle \vec{p}, \vec{q} \rangle =
-p\,q $ if $\vec{p} \perp \vec{q}$. An isotropic average of the
alignment product of two 2-dimensional vectors leads to a zero signal,
since the positive (aligned) and negative (orthogonal) contributions
cancel each other exactly. Therefore our requirement
No. \ref{cond:nonull} for suitable statistics can be guaranteed by the
usage of the alignment product.

We then define the alignment statistics of the vector fields 
$\vec{p}(\vx)= \vec{\nabla}\,$RM$(\vx)$ and
$\vec{q}=\vec{\nabla}\,$PA$_0(\vx)$ as
\begin{equation}
\label{eq:asdef}
A = A[\vec{p},\vec{q}] = \frac{\int\!d^2x\, \langle \vec{p}(\vx),
\vec{q}(\vx) \rangle}{\int \!d^2x\, |\vec{p}(\vx)|\,|
\vec{q}(\vx) |}\,,
\end{equation}
which has all required properties: 

{\bf No. \ref{cond:nofunc}}: The statistic does not depend on any
global relation between PA$_0$ and RM. This can be demonstrated by
changing a potential functional dependence using non-linear data
transformations. Any non-pathological\footnote{A pathological
transformation would e.g. split the RM or PA value range into tiny
intervals, and randomly exchange them or map them all onto the same
interval.}, piecewise continuous and piecewise monotonic pair of
transformations $\RM^* = S(\RM)$ and $\PA_0^* = T(\PA_0)$ do not
destroy the alignment signal due to the identity
\begin{equation}
\langle \vec{\nabla} {\RM}^*, \vec{\nabla} {\PA_0}^* \rangle = \left|
S'(\RM)\, T'(\PA_0) \right| \langle \vec{\nabla} \RM, \vec{\nabla}
\PA_0 \rangle.
\end{equation}
Inserted in Eq. \ref{eq:asdef} one finds that the weights of the
different contributions to the alignment signal might be changed by
the transformation, but except for pathological cases any existing
alignment signal survives the transformation and no spurious signal is
produced in the case of the null hypothesis of uncorrelated maps.

{\bf No. \ref{cond:nonull} \& \ref{cond:unique}}: A simple calculation
shows that the expectation value for $A$ is zero for independent maps
and it is unity for aligned maps. For illustration, the simulated pair
of independent maps has $A = -0.03$, which can be regarded as a
null-experiment testing our statistic with a case where RB's statistic
incorrectly detects co-alignment. The simulated pair of co-aligned
maps has $A = 0.89$ illustrating $A$'s ability to detect
correlations. Property No. \ref{cond:unique} implies requirement
No. \ref{cond:nonull}.

{\bf No. \ref{cond:analytic}}: From the discussion so far it should be
obvious that many essential properties of the alignment statistics can
be derived analytically. However, as an additional useful example we
estimate the effect of a small amount of noise present in both
maps. Each noise component is assumed to be uncorrelated with RM, to
PA$_0$, and also to the other noise component. We find for small noise
levels
\begin{equation}
\label{eq:noiseapprox}
A[\vec{p} + \delta \vec{p},\vec{q} + \delta \vec{q}] 
\approx \frac{A[\vec{p},\vec{q}]}{( 1 + {\overline{\delta
p^2}}/{\overline{p^2}})\,(1 + {\overline{\delta q^2}}/{\overline{q^2}})},
\label{eq:Anoisy}
\end{equation}
where the bar denotes the statistical average. This relation holds
only approximatively, since the non-linearity of $A$ prevents exact
estimates without specifying the full probability distribution of the
fluctuations. As can be seen from Eq. \ref{eq:noiseapprox},
uncorrelated noise reduces the alignment signal, but does not produce
a spurious alignment signal, in contrast to correlated noise, which
usually does.

{\bf No. \ref{cond:monotonic}}: If the RM and PA$_0$ maps are enlarged
by an additional region with the same statistical properties the
expectation value of $A$ is unchanged, due to the averaging property
of Eq. \ref{eq:asdef}, and fluctuations in $A$ decrease due to the
central limit theorem.

\section{Detecting map-making artifacts\label{sec:mapartefacts}}

Before the gradient alignment statistic can be applied to real
datasets, it has to be noted that it is very sensitive to correlations
on small scales, since it is a gradient square statistic. Observed RM
and PA$_0$ maps will always have some correlated fluctuations on small
scales, since they are both derived from the same set of radio maps,
so that any imperfection in the map making process leads to correlated
fluctuations in both maps. Therefore the strength of signal
contamination by such correlated errors has to be estimated, and -- if
possible -- minimized before any reliable statement about possible
intrinsic correlations of PA$_0$ and RM  can be made. This point can
not be overemphasized!

The level of expected correlated noise is investigated in Appendix
\ref{app:correlatednoise}. Fortunately, the noise correlation is of
known functional shape, which is a linear anti-correlation of the
$\PA_0$ and RM errors. This allows us to detect this noise via a {\it
gradient vector-product statistics}.
\begin{equation}
\label{eq:Vsdef}
V = V[\vec{p},\vec{q}] = \frac{\int\!d^2x\,\, \vec{p}(\vx)
\cdot \vec{q}(\vx) }{\int \!d^2x\, |\vec{p}(\vx)|\,|
\vec{q}(\vx) |}\,.
\end{equation}
This statistic is insensitive to any existing astrophysical
PA$_0$-RM correlation, since the latter should produce parallel and
anti-parallel pairs of gradient vectors with equal frequency, thereby
leading to $V=0$. A map pair without any astrophysical signal, which
was constructed from a set of independent random PA maps, will give
$V= r \gtrsim -1$, where $-1\le r\le 0$ is the correlation coefficient
of the noise calculated in Appendix \ref{app:correlatednoise}. The
statistic $V$ as a test for correlated noise fulfills therefore
requirements similar to the ones formulated for $A$, with the only
difference that a requirement like No. \ref{cond:nofunc} is not
necessary, since the functional shape of noise correlations is
known. We expect that correlated noise leads to a spurious PA$_0$-RM
co-alignment signal of the order $A \approx |V|$, since for perfectly
anti-parallel gradients these quantities are identical. Therefore, we
propose to use the quantity $A + V$ as a suitable statistic to search
for a source intrinsic PA$_0$-RM correlation: The spurious signal in
$A$ caused by the map making process should be roughly compensated by
the negative value of $V$, whereas any astrophysical co-alignment
PA$_0$-RM signal only affects $A$, and not $V$, since there should be
no preference between parallel and anti-parallel RM and PA$_0$
gradients, leading to cancellation in the scalar-product average of
$V$.

We want to apply our statistics to the RM and PA$_0$ maps of PKS~1246-410
from Taylor, Fabian \& Allen (2002), Cygnus~A from Dreher et al. (1987) and
Perley \& Carilli (1996), Hydra~A from Taylor \& Perley (1993), and 3C465 from
Eilek \& Owen (2002). Ignoring our considerations about correlated noise, an
application of our alignment statistic reveals co-spatial structures in our
set of observed maps. We find $A = 0.34,$ $0.56,$ $0.64,$ and
$0.8$,\footnote{These values are anonymized by numerical ordering. As long as
the signal is dominated by map imperfections (correlated noise, artifacts) we
refrain from specifying which map gives which $A$ and $V$ values, since this
could be regarded as a ranking of the quality of the maps. We think this would
be improper, since the observations leading to these maps were driven by other
scientific questions than we are investigating. Therefore the maps can not be
expected to be optimal for our purpose.} which have to be regarded as
significant co-alignment signals. If we use the uncorrected PA maps instead of
PA$_0$, we get only $A = 0.24,$ $ 0.35,$ and $ 0.36$ for the three sources for
which we have PA maps. These lower values indicates that the correlation
signal is mostly due to noise in the RM maps, which imprints itself to the
PA$_0$ map during the PA to PA$_0 = $PA$\, - $RM$\,\lambda^2$ correction. This
is verified by the gradient vector-product statistics, which leads to $V =
-0.32,$ $-0.36,$ $-0.58,$ and $-0.69$ for our set of maps, which clearly
reveals a preferred anti-correlation of RM and PA$_0$ fluctuations.  Visual
evidence for this can be seen in the upper middle panel of
Fig. \ref{fig:scatter}, where diagonal stripes decreasing from left to right
exhibit the presence of such anti-correlated noise.  As argued above, any real
astrophysical PA$_0$-RM correlation should be best detectable in $A+V$, for
which we get values of $A+V = -0.02,$ $-0.02,$ $0.11,$ and 0.32. This
indicates that at most in one case there could be an astrophysical
correlation, however, its significance has to be investigated since we cannot
always expect exact cancellation of $V$ and $A$ for correlated noise.

In order to suppress the spurious signal, which should be mostly
located on small spatial scales of the order of the observational
beam, in the following analysis we smooth the RM maps with a Gaussian, leaving
hopefully only the astrophysical signal carrying large-scale
fluctuations.\footnote{Smoothing of a circular quantity like a PA$_0$
angle is not uniquely defined.}

\section{Results\label{sec:results}}

Applying our statistics, we get for PKS1246-410 $A = 0.07$,
$V= -0.17$, $A+V = -0.10$ (after smoothing to a $4''$ FWHM beam), for
Cygnus A we get $A = 0.07$, $V = -0.07$, $A+V = 0.00$ (FWHM $= 2''$),
for Hydra A $A = 0.05$, $V = -0.03$, $A+V = 0.02$ (FWHM $= 1''$), and
for 3C465 $A = 0.13$, $V = 0.03$, $A+V = 0.16$ (FWHM $=9.5''$). Only
3C465 shows a marginal signature of source intrinsic co-alignment.
However, its map size in terms of resolution elements is significantly
smaller than e.g. the maps of Hydra~A and Cygnus~A, therefore a larger
statistical variance of the alignment measurement is plausible.

In contrast to the observed maps, our synthetic maps are free of
correlated noise. This is clearly revealed by our statistics, which
give for the independent pair of maps $A = 0.06$, $V = 0.05$, $A+V =
0.11$, and for the co-aligned pair of maps $A = 0.89$, $V = -0.03$,
$A+V = 0.86$. This illustrates the ability of our statistics to
discriminate between spurious and astrophysical signals.

In order to have an estimate of the statistical uncertainties, we also
apply our approach to the swapped RM map of PKS~1246-410. We get for
the PA$_{0\,{\rm East}}$-RM$_{\rm West}$ comparison $A = -0.02$, $V=
-0.19$, $A+V = -0.21$, and for the PA$_{0\,{\rm West}}$-RM$_{\rm
East}$ comparison $A = 0.10$, $V= -0.30$, $A+V = -0.20$. This
indicates that the statistical error is of the order $\delta A \sim
0.1$ and $\delta V \sim 0.2$ for this dataset, taking into account
that the swapped maps have less corresponding pixel pairs compared to
the original pair of maps.

We note that an inspection by eye reveals several sharp steps in the
RM map of PKS~1246-410, which are on length-scales below the beam size
and therefore very likely map making artifacts (see Appendix
\ref{app:correlatednoise}). Thus it is uncertain if this dataset has
sufficiently high signal-to-noise to make it suitable for a PA$_0$-RM
alignment analysis.

In summary, we do not find any evidence for a significant large-scale
co-alignment of PA$_0$ and RM maps of polarized radio sources in
galaxy clusters. 

\section{Summary\label{sec:summary}}

We investigated if there is evidence for co-aligned structures in RM
and PA$_0$ maps of extended radio sources in galaxy clusters as
claimed by RB in order to argue for source-local RM generating
magnetic fields.  

First, we have demonstrated that the null-experiment performed by RB
was poorly designed for testing the correlation between PA$_0$ and RM
in PKS~1246-410. The lack of phase coherence in the simulated data
resulted in less clustering in the simulated PA$_0$-RM scatter plots
compared to the observational scatter plots. Using independent, patchy
distributions of PA$_0$ and RM we show that the correlations due to
the mutual overlap of the RM and PA$_0$ patches produce horizontal and
vertical chains of clusters as seen in the PKS~1246-410 data, whereas
co-aligned PA$_0$ and RM patches produce diagonal stripes. The
observed clustering therefore favors independent PA$_0$ and RM maps as
expected from foreground intracluster magnetic fields.

Second, we introduce and apply a novel {\it gradient alignment
statistic} $A$.  This statistic reveals PA$_0$ and RM correlations
regardless of whether they are source intrinsic or due to artifacts in
the observation or RM map making process.  Applying this statistic to
a number of radio galaxies (PKS~1246-410, Cygnus~A, Hydra~A, and
3C465) does not reveal any significant large-scale co-alignment of
PA$_0$ and RM maps. We find significant small-scale co-alignment in
all observed map pairs, but we are able to fully identify these with
map making artifacts by another new suitable statistical test, the
{\it gradient vector product statistic} $V$.  Thus, we introduced two
new tools to analyze data of Faraday rotation studies of extended
radio sources.  They are powerful in revealing and discriminating
observational or map making artifacts (by $V$), and source-intrinsic
PA$_0$-RM correlations (by $A+V$), both indicators of potential
problems for RM based ICM magnetic field estimates.

Future, sensitive searches for potential, weak source-intrinsic
PA$_0$-RM correlations with our, or similar statistics would require
observational datasets with a much higher signal to noise ratio, and a
very well defined observational ({\it dirty}) beam. We add that such
datasets would also be crucial for detailed measurements of the
magnetic power spectra of the ICM, as proposed by En{\ss}lin \& Vogt
(2003).

We conclude that the observed RM signals of radio galaxies embedded in
galaxy clusters seems to be dominated by $\sim 1-10 \mu$G ICM magnetic
fields in accordance with independent evidence.

\acknowledgments We thank Chris Carilli for the data of Cygnus A and
for his suggestion of exchanging areas of the RM map to get
uncorrelated maps with conserved higher order statistics.  Data for
3C465 was kindly provided by Jean Eilek and Frazer Owen. We
acknowledge lively discussion with Larry Rudnick. The presentation
benefited strongly from comments of two anonymous referees, from Larry
Rudnick and Katherine Blundell, and from Matthias Bartelmann.

\appendix

\section{Parameters used for the artificial RM and PA$_0$ maps
\label{app:par}} 

The parameters to generate the artificial RM and $\PA_0$ maps are:
$N=22$ for both the RM and the PA$_0$ maps. The final map size is
$N_{\rm pxl} \times N_{\rm pxl} = 50\times50$ pixels. The seed points
are distributed by assigning
$\vx^{_{_{\RM/\PA}}}_i = ([-1.25,1.25],[-1.25,1.25])\,{\rm L}$, 
where $[a,b]$ denotes a random number, uniformly drawn from the
interval from $a$ to $b$, and $L$ denotes some arbitrary length unit
so that the final map has coordinates running from $-L$ to
$+L$. Further,
$\RM_i = [-1600,1600]\,[0,1]^{1/4}\,{\rm rad\, m^{-2}}$,
$\vec{\nabla} \RM_i = 320\,[0, 1]^{1/2} \, (\cos\theta^\RM_i,
		\sin\theta^\RM_i) \, {\rm rad\,m^{-2}\,L^{-1}}$,
		with
$\theta^\RM_i = [0,2\,\pi]$, and 
$\sigma_\RM(\vx) = [-48,48]\,{\rm rad\, m^{-2}}$.  Similarly, 
$\PA_i = [-90,90]\,{\rm deg}$,
$\vec{\nabla} \PA_i = 18\,[0, 1]^{1/2} \, (\cos\theta^\PA_i,
		\sin\theta^\PA_i) \, {\rm deg\,L^{-1}}$, with
$\theta^\PA_i = [0,2\,\pi]$, and 
$\sigma_\PA(\vx) = [-2.7,2.7]\,{\rm deg}$. Terms like $[0,
1]^{\alpha}$ lead to non-uniform distributions between 0 and 1,
which for $0<\alpha <1$ are biased towards smaller values.

\section{Avoidable and unavoidable correlated noise in RM and PA$_0$
maps\label{app:correlatednoise}} 

RM and PA$_0$ maps are generated from the same set of radio maps being
subject to observational noise. Therefore the noise of the RM and
PA$_0$ maps should be expected to be correlated. In order to
understand this correlation one has to investigate the map making
procedure.

In order to derive the RM maps, one requires a number $n$ of PA maps at
different wavelength $\lambda_i$, which we denote by $\PA_i(\vx)$. The
first crucial step is to solve the so called `n-$\pi$' ambiguity, which
arises from the fact that the PA is only defined modulo $\pi$,
whereas the determination of RM requires absolute PA values, since
\begin{equation}
\PA_i(\vx) = \PA_0(\vx) + \RM(\vx)\,\lambda_i^2
\end{equation}
allows $\PA_i$ values which deviate more than $\pi$ from $\PA_0$. Any
mistake in solving this ambiguity leads to strong step-function like
artifacts in RM, and correlated with these, steps in PA$_0$ maps. The
impact of such steps on any sensitive correlation statistics can be
disastrous. Fortunately, such n-$\pi$ ambiguities can
be strongly suppressed by using map-global algorithms to assign
absolute PAs (Dolag et al., in prep.).

However, even if the $n$-$\pi$ ambiguity is solved, observational
noise $\delta\PA_i$ in the individual maps, which we assume to be
independent from map to map, leads to correlated noise in RM and
PA$_0$, as we see in the following analysis.  Typical RM mapping algorithms use
a $\chi^2$ statistic, giving
\begin{equation}
\RM^{\rm obs} = ({\overline{\lambda^2\,\PA} -
\overline{\lambda^2}\;\overline{\PA}})/({\overline{\lambda^4} -
\overline{\lambda^2}^2}) \;, \;\;\mbox{and}\;\; \PA_0^{\rm obs} =
\overline{\PA} - \RM\, \overline{\lambda^2}\,,
\end{equation}
where we used $\overline{Y} = \sum_{i=1}^{n}\, Y_i/n$ to denote an
average of a quantity $Y$ over the different observed wavelengths. For
simplicity only, we assume similar noise levels in all maps: $\langle
\delta\PA_i^2 \rangle = \langle \delta\PA^2 \rangle\,$.  Here the
brackets denote the statistical average and should not be confused
with the alignment product. One finds for the errors $\delta \RM =
\RM^{\rm obs} - \RM$ and $\delta \PA_0 = \PA_0^{\rm obs} - \PA_0$
\begin{equation}
\langle \delta \RM^2 \rangle = {\langle \delta\PA^2
\rangle}/({n\,(\overline{\lambda^4} - \overline{\lambda^2}^2)})\,,\;
\;
\langle \delta \PA_0^2 \rangle = \overline{\lambda^4}\,\langle \delta
\RM^2 \rangle\,, \;\;\mbox{and}\;\;
\langle \delta \RM\,\delta \PA_0 \rangle = - \overline{\lambda^2}\,\langle
\delta \RM^2 \rangle \,,
\end{equation}
which implies an anti-correlation between $\PA_0$ and RM noise, with a
correlation coefficient of
\begin{equation}
r = {\langle \delta \RM\,\delta \PA_0 \rangle}/{\sqrt{ \langle \delta
\RM^2 \rangle \, \langle \delta \RM^2 \rangle }} = -
{\overline{\lambda^2}}/{\sqrt{\overline{\lambda^4}}}\,.
\end{equation}
For the set of frequencies used to derive the RM map of PKS~1246-410
one finds $r = -0.89$, therefore we expect the noise in the RM and
PA$_0$ maps of PKS~1246-410 to be highly anti-correlated. The scalar
product of the gradients, which enter the gradient scalar-product
statistic $V$, have expectation values of
\begin{equation}
\langle \vec{\nabla} \RM^{\rm obs} \cdot \vec{\nabla} \PA_0^{\rm obs}
\rangle = \langle \vec{\nabla} \delta \RM \cdot \vec{\nabla} \delta
\PA_0 \rangle = - \langle |\vec{\nabla}\delta\PA|^2 \rangle
\overline{\lambda^2}/(n(\overline{\lambda^4} -
\overline{\lambda^2}^2)).
\end{equation}
Similarly, one finds
\begin{equation}
\langle |\vec{\nabla} \delta \RM|^2 \rangle = \langle
|\vec{\nabla}\delta\PA|^2 \rangle/(n(\overline{\lambda^4} -
\overline{\lambda^2}^2))\;,\;\; \mbox{and}\;\;
\langle |\vec{\nabla} \delta \PA_0|^2 \rangle = \langle
|\vec{\nabla}\delta\PA|^2
\rangle\,\overline{\lambda^4}/(n(\overline{\lambda^4} -
\overline{\lambda^2}^2))
\end{equation}
so that for noise dominated maps there is a strong signal in the
gradient scalar-product statistic $V = r = -
{\overline{\lambda^2}}/{\sqrt{\overline{\lambda^4}}}$. In the case of
non-constant astrophysical RM and PA$_0$ values, $|V|$ will be
smaller. However, since noise is usually strongest on the smallest
scales, the gradient of the noise can easily be stronger than the
gradient of the astrophysical signal, leading to $V \sim r$.

\clearpage

\end{document}